# Modeling the Resilience of Large and Evolving Systems

MOHAMED KAANICHE[1], PAOLO LOLLINI[2], ANDREA BONDAVALLI[2], and KARAMA KANOUN[1]

[1]*LAAS-CNRS, University of Toulouse, 7 av. Colonel Roche, 31077 Toulouse - FRANCE*
[2]*University of Florence-DSI, viale Morgagni 65, I-50134, Florence - ITALY*

**Abstract**: This paper summarizes the state of knowledge and ongoing research on methods and techniques for resilience evaluation, taking into account the resilience-scaling challenges and properties related to the ubiquitous computerized systems. We mainly focus on quantitative evaluation approaches and, in particular, on model-based evaluation techniques that are commonly used to evaluate and compare, from the dependability point of view, different architecture alternatives at the design stage. We outline some of the main modeling techniques aiming at mastering the largeness of analytical dependability models at the construction level. Actually, addressing the model largeness problem is important with respect to the investigation of the scalability of current techniques to meet the complexity challenges of ubiquitous systems. Finally we present two case studies in which some of the presented techniques are applied for modeling web services and General Packet Radio Service (GPRS) mobile telephone networks, as prominent examples of large and evolving systems.

## 1. Introduction

The main objectives of dependability and resilience evaluation activities are to support design decision-making and to assess the level of confidence that can be placed on the target systems with respect to their ability to fulfill their missions at the desired dependability and security level. Resilience evaluation can be qualitative or quantitative, and the evaluation techniques can be based on analytical modeling, simulation, experimental evaluation or judgments. Models can be used at the early development stages to describe various alternative architectures of the system at different abstraction levels, and to analyze the impact on the system behavior of different threat assumptions, error detection and recovery strategies, maintenance policies, etc.

This paper addresses model-based dependability evaluation approaches focusing on the techniques proposed in the literature for mastering models largeness and complexity. To master complexity when evaluating system dependability, a modeling methodology is needed so that only the relevant system aspects need to be detailed, allowing numerical results to be effectively computable. The complexity of models depends on the dependability measures to be evaluated, the modeling level of detail, and the stochastic dependencies among the components. State-space models, in particular homogeneous Markov chains, are commonly used for dependability modeling of computing systems. They are able to capture various functional and stochastic dependencies among components and allow evaluation of various measures related to dependability and performance (i.e., performability measures) based on the same model, when a reward structure is associated to them. Unfortunately not all the existing systems and their features can be captured properly by Markov processes; in some cases more general processes (e.g., semi-Markov, Markov Regenerative or even more non-Markovian processes) must be used. When dealing with such processes, complex and costly analytical solution techniques may exist, and otherwise simulation is the approach used to solve the models thus providing only estimates of the measures of interest. As an alternative one can approximate an underlying non-Markovian process with a Markov one, and thus represent a non-exponential transition with an appropriate set of exponential ones (Phased-Type approach). The price to pay following this approach is a significant increase in size, in terms of number of states of the resulting model.

Significant progress has been made in addressing the challenges raised by the large size of models, at the model construction and model solution levels, using a combination of techniques that can be categorized with respect to their purpose (largeness avoidance, largeness tolerance). Largeness avoidance techniques try to reduce the size of the models using many different approaches. Largeness tolerance techniques try to optimize the generation and processing of the models through the use of a) systematic rules to support the elaboration of the models and b) space and time efficient algorithms to optimize the state space generation, storage and exploration. It is important to note that these categories of techniques are complementary and both are needed, at the model construction and model solution levels, when detailed and large dependability models need to be generated and processed to evaluate metrics characterizing the resilience of real life systems.

In this paper we focus on the modeling approaches aimed at mastering complexity at the model construction level. We present in Section 2 two classes of structured techniques for a modular model construction. Two examples of case studies are presented in Section 3: the first one addresses the modeling of web based services and systems, and the second one deals with the assessment of a mobile telephone system. Finally, Section 4 concludes the paper and discusses some research gaps and directions that still need to be investigated to address the challenges raised by large and evolving ubiquitous systems.

## 2. Model construction techniques

This section presents two complementary techniques that can be applied in the model construction phase to cope with model largeness, grouping them in composition approaches (Section 2.1) and decomposition/aggregation approaches (Section 2.2). We focus on analytical models using Markov chains, stochastic Petri nets and their extensions.

### 2.1. Composition approaches

This class groups those techniques that build the system model as a modular composition of simpler sub-models that are then solved as a whole. Most of the works belonging to this class define the rules to be used to construct and interconnect the sub-models. They provide an easy way to describe the behavior of systems having a high degree of dependency between subcomponents. These dependencies can be exploited to manage the model complexity creating, for example, smaller, equivalent representations. In the following, we firstly outline briefly some general composition approaches based on Stochastic Petri nets and their extensions. Then we focus on specific studies dedicated to the dependability modeling and evaluation of complex real-life systems.

### 2.1.1. Basic approaches based on Stochastic Petri nets and their extensions

Research on process algebra [1] has inspired efforts to introduce compositionality into Petri nets. Composition of Petri Net (PN) consists in constructing PN models from a set of building blocks by applying suitable operators of places and/or transition superposition. Initially investigated for no timed models, composition approaches have been then explored for stochastic extensions of Petri nets. For example, [2] explored composition in the context of Stochastic Petri Nets (SPNs) and [3] proposed a systematic compositional approach to the construction of parallel hardware-software models using Generalized Stochastic Petri Nets (GSPNs). The GSPN composition rules are based on the concept of matching labels associated with transitions and places of a GSPN, and the superposition of transitions (places) with matching labels, each one belonging to a different GSPN.

Composition operators have been defined in [4, 5] for the generation of Stochastic Well-formed Nets (SWN) (i.e., GSPNs permitting the identification of symmetry by means of a symmetric reachability graph) from its components. These operators preserve the functional structure of the model and support several types of communication between components. This approach is intended to support the modeling of distributed and parallel systems where both synchronous and asynchronous communications are required. However, it addresses only the class of systems that can be modeled by SWN.

Another example of composition operators is used in the context of Stochastic Activity Networks (SAN). In [6], two composition operators are defined (*Join* and *Replicate*) to compose system models based on SANs. The Join operator takes as input a set of sub-models and some shared places belonging to different sub-models, and provides as output a new model that comprehends all the joined sub-models elements (places, arcs, activities) but with the shared places merged in a unique one. The Replicate operator combines multiple identical copies of a sub-model, which are called replicas, sharing some selected places. [7] introduces a graph composition approach that extends the replicate/join formalism and also combines models by sharing a portion of the state of each sub-model, reducing the total state-space size. Contrarily to the join/replicate formalism that requires the use of a special operation, the graph composition detects all the symmetries exposed at the composition level and uses them to reduce the underlying state space.

**2.1.2. Compositional approaches with dependencies modeling**

The composition techniques discussed in subsection 2.1.1 are very helpful to cope with the models complexity, in particular when the models exhibit symmetries. However, they are not sufficient in particular when the modeled systems exhibit various dependencies that need to be explicitly described in the dependability models. These dependencies may result from functional or structural interactions between the components or from interactions due to global system fault tolerance, reconfiguration and maintenance strategies. Various modeling approaches have been proposed to facilitate the construction of large dependability models taking into account such dependencies. Examples of such modeling approaches are briefly discussed in the sequel.

The block modeling approach defined in [8] provides a generic framework for the dependability modeling of hardware and software fault-tolerant systems based on GSPNs. The proposed approach is modular: generic GSPN submodels called block nets are defined to describe the behavior of the system components and of the interactions between them. The system model is obtained by composition of these GSPNs. Composition rules are defined and formalized through the identification of the interfaces between the component and interaction block nets. In addition to modularity, the formalism brings flexibility and re-usability thereby allowing for easy sensitivity analyses with respect to the assumptions that could be made about the behavior of the components and the resulting interactions. The main advantage of this approach lies in its efficiency for modeling several alternatives for the same system as illustrated for example in [9].

The efficiency of the block modeling approach can be further improved by using an incremental and iterative approach for the construction and validation of the models as suggested in [10]. At the initial iteration, the behavior of the system is described taking into account the failures and recovery actions of only one selected component, assuming that the others are in an operational nominal state. Dependencies between components are taken into account progressively at the following iterations. At each iteration a new component is added and the GSPN model is updated by taking into account the impact of the additional assumptions on the behavior of the components that have been already included in the model. Similarly to the block modeling approach, sub-models are defined for describing the components behaviors and specific rules and guidelines are defined for interconnecting the submodels. This approach has been successfully used to model the dependability of the f the French air traffic control computing system [11].

An iterative dependability modeling approach has been also proposed in [12] where the construction and validation of the GSPN dependability model is carried out progressively following the system development refinement process, to facilitate the integration of dependability modeling activities in the system engineering process. Three main steps are distinguished. The first step is dedicated to the construction of a functional-level model describing the system functions, their states and their interdependencies. In the second step, the functional level model is transformed into a high-level dependability model based on the knowledge of the system's structure. A model is generated for each pre-selected candidate architecture. The third step is dedicated to the refinement of the high-level dependability model into a detailed dependability model for each selected architecture. Formal rules are defined to make the successive model transformations and refinements as systematic as possible taking into account three complementary aspects: i) component decomposition, ii) state/event fine-tuning, and iii) stochastic distribution adjustments. This approach allows the integration of various dependencies at the right level of abstraction: functional dependency, structural dependency and those induced by non-exponential distributions. A case study is described in [13].

Actually, the approach presented in [12] can be seen as a special case of the more general class of techniques based on layered and multi-level modeling methods, where the modeled system is structured into different levels corresponding to different abstraction layers, with a model associated to each level. Various techniques based on this idea have been developed, see e.g., [3, 14-21]. As an example, we can mention the multilevel modeling approach proposed in [14] for evaluating the user perceived availability of web-based applications that is discussed in Section 3.1.

Generally, multilevel modeling approaches rely on the hierarchical composition and solution of the submodels corresponding to the different abstraction levels. Different techniques can be used to describe the submodels and to combine their results (combinatorial models, state-based models). The selection of the right technique depends on the kind of dependencies between the elements of the corresponding submodels and on the quantitative measures to be evaluated. It is noteworthy that hierarchical modeling approaches combining different types of models can be supported by automatic tools, e.g., SHARPE [22, 23], Draw-Net [24]. These approaches belong to the more general class of model decomposition and aggregation approaches presented in Section 2.2.

## 2.2. Decomposition and aggregation approaches

The modeling approaches discussed in this section belong to largeness avoidance techniques that try to circumvent the generation of large models using model decomposition and aggregation of the partial results. The basic idea is the following: the overall model is decoupled into simpler and more tractable sub-models, and the measures obtained from the solution of the sub-models are then aggregated to compute those concerning the overall model. The decomposition and aggregation techniques depend on the type of measures to be evaluated (steady-state or transient) and on the modeling formalism. Generally approximate solutions are provided for the composition of the results derived from the submodels. In the following, we present different examples of decomposition/aggregation approaches proposed in the literature.

A decomposition and aggregation theory for steady state analysis of general continuous time Markov Chains has been proposed in [25]. The quality of the approximation is related to the degree of coupling between the blocks into which the Markov chain matrix is decomposed. In [26] the authors present an extension of this technique specifically addressed to the transient analysis of large stiff Markov chains, where stiffness is caused by the simultaneous presence of "fast" and "slow" rates in the transition rate matrix.

Time-scale based decomposition approaches have been applied to Non-Markovian stochastic systems in [27], and to GSPN models of systems containing activities whose durations differ by several orders of magnitude in [28]. For example, in [28] the GSPN model is decomposed into a hierarchical sequence of aggregated sub-nets each of which is characterized by a certain time scale. Then these smaller sub-nets are solved in isolation, and their solutions are combined to get the solution of the whole system. The aggregation at each level is done by assuming that the transitions included in the lower level are immediate transitions. At each level of the hierarchy, the current marking of an aggregated sub-net determines the number of tokens in the sub-net at the lower level, which are then analyzed to determine the rate of transitions in the aggregated sub-net.

Another interesting extension of the decomposability theory presented in [25] is the decomposition approach for the solution of large stochastic reward net models proposed in [29]. The overall model consists of a set of submodels whose interactions are described by an import graph: each node of the graph corresponds to a parameterized stochastic reward net submodel and an arc from submodel A to submodel B corresponds to a parameter value that B must receive from A. It is shown that three quantities are sufficient for intercommunication between the defined subnets: the probability that a subnet is in a state satisfying a given condition, the average time a given condition remains satisfied, and the expected time until the subnet satisfies a given condition.

The decomposition approach in [30] is based on a new set of connection formalisms that reduce the state-space size and solution time by identifying submodels that are not affected by the rest of the model, and solving them separately. The result from each solved submodel is then used in the solution of the rest of the model. The authors develop four abstractions that can be used to make connection models, and they involve passing a continuous-time random process, a discrete-time random process, a random variable, and an average value between the models. When these abstractions are applied, each submodel should have a smaller state space and fewer time scales than the complete model.

Decomposition approaches are also relevant for the modeling of multiple-phased systems having different operational phases and regimes with different configurations, behaviors, and dependability characteristics. In the literature, several approaches have been proposed for the analytical dependability modeling of Phased Mission Systems (PMS), all based on a hierarchical structure of the models [31-33]. PMS are characterized by a sequence of phases in which the system configuration can change during operations. The existence of phases is a consequence of: i) diverse tasks to be performed, and ii) diverse environmental conditions, in different periods of system lifetime.

In [34], the model of a PMS is seen as composed of two logically separate Petri nets: the System Net, representing the system (its components, their interactions and their failure/repair behavior) as a GSPN, and the Phase Net, a deterministic and Stochastic Petri Net, representing the control part and describing the phase changes. In the System Net, a single model is built for the whole mission, characterized by a set of phases without detailing the behavior of the system inside each phase. This allows easy modeling of a variety of mission scenarios by sequencing the phases in appropriate ways. The parameter values to be used in the System Net model are obtained by solving the Phase Net models. This approach has been generalized in [35] based on Markov Regenerative Stochastic Petri Nets. The key point is that the state space of the Markov regenerative process is never generated and handled as a whole, but rather the various subordinate intra-phase processes are separately generated and solved. As a consequence, the computational complexity of the analytical solution is reduced to the one needed for the separate solution of the different phases, as demonstrated in [36].

In [37], the authors proposed a decomposition and aggregation approach that operates at the system-level, rather than the model-level. Using this approach, entities (or sub-systems) are created that can work in isolation or can interact with each other through a set of dependency relations. The relations state how the behavior of each entity affects the others. The structure, together with the notion of a phased mission, allows one to solve each submodel in isolation, and then pass results between submodels as needed. Such formulation is not domain-specific and it reduces the complexity of solving models that can be expressed in this framework. This generic decomposition/aggregation approach has been applied to study a GPRS mobile telephone infrastructure that takes into account the congestion due to service outages and its subsequent impact on user-perceived quality of service. More details about this case study are presented in Section 3.2.

## 3. Case studies

This section summarizes the results of two case studies illustrating some of the methodologies and modeling approaches discussed in the previous section. The first example concerns the dependability modeling of web-based systems and services and the second example is related to the QoS analysis of mobile telephone systems.

### 3.1. Dependability modeling of web-based systems and services

Growing usage of applications on the Internet make the issue of assessing the dependability of the delivered services as perceived by the users increasingly important. The Internet is often used for money critical applications such as online banking, stock trading, reservation processing and shopping, where the temporary interruption of service could have unacceptable consequences on the e-business. Thus, it is important for e-business service providers and developers to analyze during the architecture design phase how hardware, software and performance related failures of the infrastructure supporting the delivered services might affect the quality of service perceived by the users.

Quantitative measures characterizing user-perceived availability for web-based applications are widely recognized as highly important to evaluate the impact of failures from the business point of view. However, there is still a lack of modeling frameworks and examples illustrating how to address this issue. Indeed a significant body of work has focused on various aspects of web performance evaluation. Although many efforts have been dedicated to analyze the availability of web hosts using measurement-based techniques [38, 39], less emphasis has been put on the modeling of web service availability taking into account the impact of server node failures and performance degradations (see [40] for a review of the state-of-the art). In the following, we summarize the multi-level modeling framework proposed in [14] in order to address this gap.

Internet-based applications are generally implemented on largely distributed infrastructures, involving various types of servers such as web, application, and database servers. Three key players are typically involved in the provision of the services delivered by such applications: 1) the users (i.e., the customers), 2) the e-business provider (eBP), who implements the e-business functions invoked by the users; these functions are based on a set of services and resources that are internal to the eBP site(s) or are provided by external suppliers, and 3) the external suppliers. Considering the example of a web-based travel agency allowing the customers to plan and book trips over the web, the external suppliers correspond to the flight reservation systems (AF, KLM, …), hotel reservation systems (Sofitel, Holiday Inn, …), and car rental systems (Hertz, Avis, …) that are accessed by the travel agency application through dedicated interfaces.

Generally, the eBP has a full control of its own architecture. Therefore, a detailed dependability modeling and analysis of this architecture can be carried out to support architectural design decisions. However, only limited information is generally available to analyze the dependability of the external suppliers services. In this context, remote measurements can be used to evaluate some parameters characterizing the dependability of these services. These parameters can then be incorporated into the models describing the impact of eBP component failures and repairs on the user perceived dependability.

The discussion above shows that several issues should be taken into account when modeling the user perceived availability of Internet based applications. Due to the complexity of the target system and the difficulty to combine various types of information (users behavior, failure-recovery scenarios of the supporting infrastructure), a systematic and pragmatic approach is needed to support the construction of such dependability models. The framework presented in [14] proposed a multilevel hierarchical approach for modeling the user perceived availability of e-business applications. Modeling is done in two steps: 1) identification of the functions and services provided to the users and the resources contributing to their accomplishment, and characterize how the users interact with the application, and, 2) based on this, building of model(s) to assess the impact of component failures and repairs on the quality of service delivered to the users.

As sketched in Figure 1, four abstraction levels are distinguished, namely, *user*, *function*, *service* and *resource* levels. The highest level (user level) describes the availability of the application as perceived by the users. Intermediate levels describe the availability of functions and services provided to the users. The lowest level (resource level) describes the availability of the component systems on which functions and services are implemented. The availability measures of a given level are obtained based on the measures computed at the immediately lower level. Various techniques can be used to model each level: fault trees, reliability block diagrams, Markov chains, stochastic Petri nets, etc. The selection of the right technique mainly depends on the kinds of dependencies between the elements of the considered level and on the quantitative measures to be evaluated. This approach is illustrated in [41] using a web-based travel agency as an example. In addition detailed analytical performability models are presented in [42] to analyze the availability of web services implemented on cluster architectures taking into account explicitly: 1) the cluster architecture characteristics, considering the number of nodes in the cluster, the error recovery strategy after a node failure, as well as the reliability of cluster nodes, 2) the traffic model, describing the web traffic characteristics (i.e., the access patterns), and 3) various causes of service unavailability including requests loss due to buffer overflow or node failures.

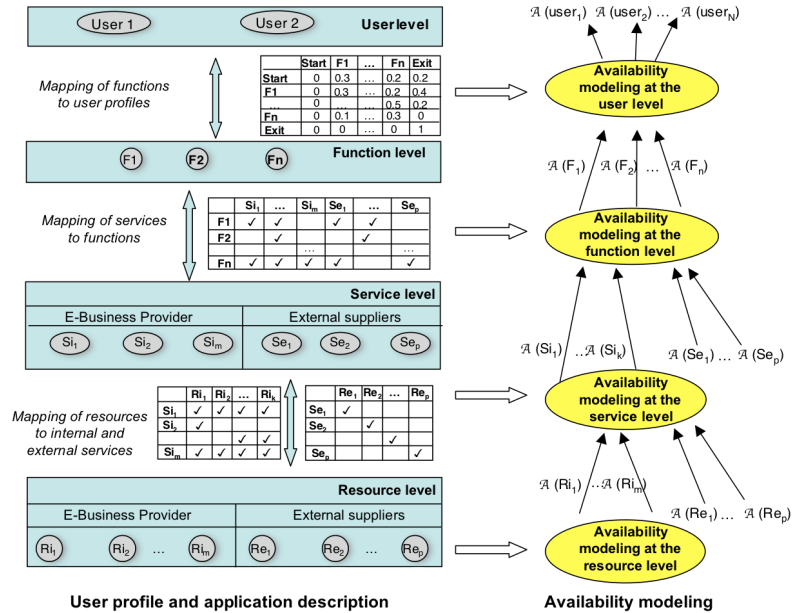

**Fig. 1: Hierarchical availability modeling framework**

For the sake of illustration, in the web-based travel agency case study presented in [41], we have evaluated the user-perceived unavailability considering two different user profiles and various alternative fault tolerant architectures for the implementation of the application and internal services. The first profile corresponds to users who are mainly seeking for information without a buying intention, whereas in the second profile the percentage of transactions that end up with the booking and payment of a trip is three times higher. It is shown that such a difference could have a significant impact on the user perceived unavailability (e.g., 173 hours downtime per year for the first profile compared to 190 for the second profile for one on the investigated scenarios).

Another facet of the web-services analyses has been studied in [43] considering a business model workflow that fits into the class of systems composed of multiple phases. It is shown that modeling methodologies and tools for dependability analysis of Multiple Phased Systems [44] can be applied to provide a useful support to service providers in choosing the most appropriate service alternatives to build their own composite services.

### 4.2. QoS analysis of Mobile Telephone Systems

Mobile telephone systems are prominent representatives of ubiquitous systems and their dependability, especially in terms of availability, has been extensively analyzed in the past literature. Besides the mere estimation of system availability, in recent years several works have been presented focusing on the QoS analysis from the users' perspective, analyzing how the user's perceived QoS is affected by outages, congestion events and by the application of appropriate congestion treatment techniques. The high system complexity in terms of largeness, heterogeneity and dynamicity has been attacked through the introduction of appropriate modeling methodologies based on composition as well as decomposition approaches. In this section we focus on works addressing mobile telephone systems based on General Packet Radio Service (GPRS) technology.

The work presented in [45, 46] contributes to the analysis of a GPRS network by providing a hierarchical/composition modeling approach to understand the effects of outage periods on the service provision, adopting the modeling capabilities of Stochastic Activity Networks. A typical GPRS configuration has been deeply analyzed and evaluated in terms of a few identified QoS indicators, namely, the number of served users per hour (and related measures) and the variation in system availability when including outages effects with respect to the bare network availability analysis. Interesting results have been observed, which can be fruitfully exploited in devising GPRS configurations adequate to maintain an acceptable QoS also in critical, overload conditions. Additionally, some sensitivity analysis has been performed by varying the frequency and duration of outages periods and for different values of network availability.

The congestion analysis of GPRS infrastructures consisting of a number of cells partially overlapping has been performed in [47], in terms of QoS indicators expressing a measure of the service availability perceived by users. When congestion is experienced by one of these cells, a family of congestion management techniques is put in place, to operate a redistribution of a number of users in the congested cell to the neighbor ones, in accordance with the overlapping areas. Since the service availability perceived by users is heavily impacted by the congestion experienced by the cells, determining appropriate values for the users to switch, so as to obtain an effective balance between congestion alleviation in the congested cell and congestion inducement in the receiving cells, is a critical aspect in this context. In order to carry on such fine-tuning activity, a decomposition/aggregation modeling methodology, appropriate to deal with the system complexity, has been defined. In particular, a top-down approach is adopted to move from the entire system description to the definition of more simple sub-models. Then, the model solution process follows a bottom-up compositional approach.

The obtained results show behavior trends very useful to make an appropriate choice of the number of users to switch, for example emphasizing that it is not always useful to switch as much users as possible from the congested cell to the other(s), since the positive effects induced in the congested cell do not always compensate the negative effects on the receiving cells. Moreover, an investigation on the amount of time that the system should be permitted to spend for its decision-making processes has been carried on.

A more general decomposition/aggregation approach based on the interactions between system subcomponents has been presented in [37]. The goal was to derive a domain-free modeling framework, not restricted to the analysis of mobile telephone systems but applicable to a more general class of systems sharing some high-level properties.

The overall modeling and solution process is depicted in Figure 2. The conceptual-level system decomposition is a combination of a *functional* and *temporal* decomposition. The system is first analyzed from a functional point of view: the overall system is decomposed in a set of interacting sub-systems, called "entities" (the numbered circles of Figure 2), each one corresponding to a critical system function with respect to the validation objectives. The entities may interact through dependency relations: entity Y depends on entity X if the behavior of Y is affected by X (in Figure 2, X→Y). Then a temporal decomposition is applied, for which the system lifetime is seen as a sequence of phases such that each phase is characterized by the same dependency relations holding among entities, and at least one dependency relation changes between two consecutive phases. At this point the structure of the obtained equivalent system (called "*phased-interacting*" system) is exploited to propose a decomposed modeling approach that also includes some capabilities for managing the system complexity. Accordingly, each entity is modeled in isolation using the most appropriate modeling formalism. Each model X* (the numbered squares of Figure 2) represents the behavior of entity X including its dependency relations. The decoupled models interact with each other through the passing of numerical results (the dashed arrows of Figure 2): an intermediate result produced by the solution of a given model is shared on a Data Base and it is used to set the value of a numerical parameter defined in another model. The model X*, together with the intermediate results provided as input, should reconstruct the original stochastic behavior of entity X. Then the decomposed solution process can be applied: in each phase, the decoupled models (with the available intermediate results) are solved in isolation using the most appropriate solution technique and following the order defined by the dependency relation graph, finally obtaining the measures of interest. The accuracy of the final measures depends on the accuracy of the intermediate results and on the capability of the decoupled models to properly reconstruct the entities' interactions.

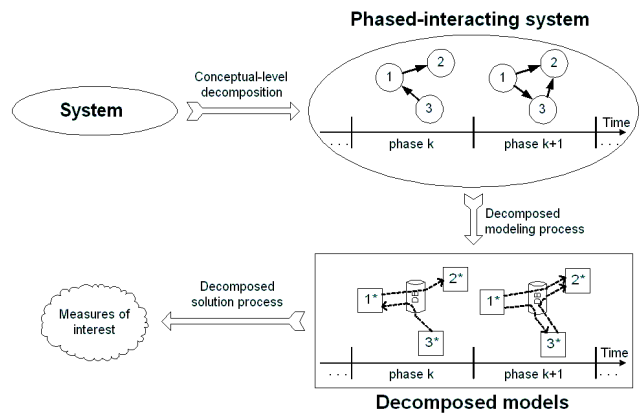

**Fig. 2: Interaction-based decomposition approach**

The sketched general modeling and solution approach has been applied to analyze the same GPRS scenario presented in [47]. The goal was to demonstrate its usage in a concrete application domain and to check for its effectiveness and the accuracy of the final measures. In the corresponding phased-interacting system, each entity corresponds to a GPRS cell, the dependency relations represent the users' switching among cells (as consequence of the application of the congestion treatment techniques), and the beginning/end of the switching procedures determines the phase transitions. With respect to the study performed in [47]: i) the accuracy of the obtained results is acceptable, ii) the computational time slightly increases (the methodology presented in [47] has been specifically developed for the selected GPRS scenario, this is why it is more efficient), but it is substantially reduced compared to solving the whole non-decomposed model.

## 3. Conclusions

The increasing scale and complexity of modern-day computing systems continues to put a premium on efficient techniques for the construction and solution of large quantitative models. In addition, these large, dynamic and evolving systems pose some new research challenges.

In this paper, we have focused on model-based analytical evaluation techniques. The complexity of the scenarios to be considered for the assessment of large and dynamic systems calls for the development of holistic evaluation approaches including complementary evaluation techniques, covering simulation, analytical modeling and experimental measurements. Mechanisms are needed to ensure the cooperation and the unified integration of these techniques, in order to provide realistic assessments of architectural solutions and of systems in their operational environments.

More generally, evaluation methods must consider metrics at an increasingly high level of abstraction, to express the impact of the computing infrastructure on an enterprise business. Of increased significance is also the use of quantitative evaluation methods to support the effective use of adaptation mechanisms prevalent in modern-day systems. Mechanisms exist to dynamically change configurations, and well-founded quantitative techniques are necessary to run such systems in a resilient way.

Besides assessing the impact accidental threats, extensions are also needed to quantify the impact of malicious threats. Clearly, there is a need for a comprehensive modeling framework that can be used to assess the impact of accidental faults as well as malicious threats in an integrated way.

**Acknowledgement**

This work has been partially supported by the European Commission (IST Projects ReSIST 0265764, HIDENETS 026979 and CRUTIAL 027513).

**References**

[1] R. Milner, *Communication and Concurrency*, Prentice Hall, 1989.
[2] P. Buchholz, "A Notion of Equivalence for Stochastic Petri Nets," *16th Int. Conf. on Application and Theory of Petri Nets,* Torino, Italy, 1995, pp. 161-180.
[3] S. Donatelli and G. Franceschinis, "The PSR methodology: integrating hardware and software models," *17th Int. Conf. on Application and Theory of Petri Nets, ICATPN '96,* Osaka, Japan, 1996, (Springer-Verlag).
[4] I. Rojas, "Compositional Construction of SWN Models," *The Computer Journal*, vol. 38, no. 7, pp. 612-621, 1996.
[5] S. Ballarini, S. Donatelli and G. Franceschinis, "Parametric Stochastic Well-Formed Nets and Compositional Modelling," in *21st International Conference on Application and Theory of Petri Nets,* Aarhus, Denmark, 2000, (Springer Verlag).
[6] J. F. Meyer and W. H. Sanders, "Specification and Construction of Performability Models," in *Int. Workshop on Performability Modeling of Computer and Communication Systems,* Mont Saint Michel, France, 1993, pp. 1-32.
[7] W. D. Obal, *Measure-Adaptive State-Space Construction Methods,* PhD, University of Arizona, 1998.
[8] K. Kanoun and M. Borrel, "Fault-Tolerant System Dependability — Explicit Modeling of Hardware and Software Component-Interactions," *IEEE Transactions on Reliability*, vol. 49, no. 4, pp. 363-376, December 2000.
[9] K. Kanoun, M. Borrel, T. Moreteveille and A. Peytavin, "Modeling the Dependability of CAUTRA, a Subset of the French Air Traffic Control System," *IEEE Transactions on Computers*, vol. 48, no. 5, pp. 528-535, 1999.
[10] N. Fota, M. Kâaniche and K. Kanoun, "Incremental Approach for Building Stochastic Petri Nets for Dependability Modeling," in *Statistical and Probabilistic Models in Reliability,* (Ionescu and Limnios, Eds.), pp. 321-335, Birkhäuser, 1999.

[11] N. Fota, M. Kâaniche and K. Kanoun, "Dependability Evaluation of an Air Traffic Control Computing System," *Performance Evaluation*, vol. 35 (3-4), 553-573, 1999.
[12] C. Betous-Almeida and K. Kanoun, "Construction and Stepwise Refinement of Dependability Models," *Performance Evaluation*, vol. 56, 277-306, 2004.
[13] C. Betous-Almeida and K. Kanoun, "Dependability modelling of Intsrumentation and Control Systems: A Comparison of Competing Architectures," *Safety Science*, vol. 42, 457-480, 2004.
[14] M. Kaâniche, K. Kanoun and M. Rabah, "Multi-level modelling approach for the availability assessment of e-business applications," *Software: Practice and Experience*, vol. 33, no. 14, pp. 1323-1341, 2003.
[15] S. Bernardi and S. Donatelli, "Building Petri Net Scenarios for Dependable Automation Systems," in *10th International Workshop for Petri Nets and Performance Models (PNPM'2003),* Urbana-Champaign, IL, USA, 2003, pp. 72-81.
[16] S. Bernardi, *Building Stochastic Petri Net models for the Verification of Complex Software Systems,* PhD, Universita di Torino, 2003.
[17] M. Rabah and K. Kanoun, "Performability evaluation of multipurpose multiprocessor systems: the "separation of concerns" approach," *IEEE transactions on Computers*, vol. 52, no. 2, pp. 223-236, 2003.
[18] A. Bondavalli, M. Nelli, L. Simoncini and G. Mongardi, "Hierarchical Modelling of Complex Control Systems: Dependability Analysis of a Railway Interlocking," *Journal of Computer Systems Science and Engineering*, 16(4), pp. 249-261, 2001.
[19] P. Lollini, A. Bondavalli and F. Di Giandomenico, "A modeling methodology for hierarchical control systems and its application," *Journal of the Brazilian Computer Society*, vol. 10, no. 3, pp. 57-69, 2005.
[20] M. Nelli, A. Bondavalli and L. Simoncini, "Dependability Modelling and Analysis of Complex Control Systems: an Application to Railway Interlocking," *European Dependable Computing Conference (EDCC-2),* Taormina, Italy, 1996, pp. 93-110, (Springer-Verlag).
[21] Y.-S. Dai, Y. Pan and X. Zou, "A Hierarchical Modeling and Analysis for Grid Service Reliability," *IEEE Trans. on Computers*, vol. 56, no. 5, pp. 681-691, 2007.
[22] R. A. Sahner and K. S. Trivedi, "Reliability modeling using SHARPE," *IEEE Transactions on Reliability*, vol. R-36, no. 2, pp. 186-193, 1987.
[23] R. A. Sahner, K. Trivedi and A. Puliafito, *Performance and Reliability Analysis of Computer Systems: An Example-based Approach Using the SHARPE Software Package*, Kluwer Academic Publisher, 1996.
[24] M. Gribaudo, D. Codetta-Raiteri and G. Franceschini, "Draw-Net, a customizable multi-formalism multi-solution tool for the quantitative evaluation of systems," *2nd International Conference on Quantitative Evaluation of Systems,* Torino, Italy, 2005.
[25] P. J. Courtois, *Decomposability - Queueing and Computer System Applications*, New York: Academic Press, 1977.
[26] A. Bobbio and K. Trivedi, "An Aggregation Technique for the Transient Analysis of Stiff Markov Chains," *IEEE Transactions on Computers*, vol. C-35, no. 9, pp. 803-814, August 1986.
[27] S. Haddad and P. Moreaux, "Approximate Analysis of Non-Markovian Stochastic Systems with Multiple Time Scale Delays," *12th Annual Meeting of the IEEE International Symposium on Modeling, Analysis, and Simulation of Computer and Telecommunication Systems (MASCOTS)* Volendam, NL, 2004.

[28] H. H. Ammar and S. M. Rezaul Islam, "Time scale decomposition of a class of generalized stochastic Petri net models," *IEEE Transactions on Software Engineering*, vol. 15, no. 6, pp. 809-820, 1989.

[29] G. Ciardo and K. S. Trivedi, "Decomposition Approach to Stochastic Reward Net Models," *Performance Evaluation*, vol. 18, no. 1, pp. 37-59, 1993.

[30] D. Daly and W. H. Sanders, "A connection formalism for the solution of large and stiff models," *34th Annual Simulation Symposium,* 2001, pp. 258-265.

[31] J. B. Dugan, "Automated Analysis of Phase-Mission Reliability," *IEEE Transaction on Reliability*, vol. 40, 45-52, 1991.

[32] M. Alam and U. M. Al-Saggaf, "Quantitative Reliability Evaluation of Repairable Phased-Mission Systems using Markov Approach," *IEEE Transaction on Reliability*, vol. 35, 498-503, 1986.

[33] A. Somani and K. Trivedi, "Phased-mission System Analysis using Boolean Algebraic Methods," in *1994 ACM SIGMETRICS conference on Measurement and modeling of computer systems,* Nashville, Tennessee, USA, 1994, pp. 98-107.

[34] I. Mura, A. Bondavalli, X. Zang and K. Trivedi, "Dependability Modelling and Evaluation of Phased Mission Systems: a DSPN Approach," *7th IFIP Int. Conference on Dependable Computing for Critical Applications (DCCA-7),* San Jose, CA, USA, 1999, (IEEE Computer Society).

[35] I. Mura and A. Bondavalli, "Markov Regenerative Stochastic Petri Nets to Model and Evaluate the Dependability of Phased Missions," *IEEE Transactions on Computers*, vol. 50, no. 12, pp. 1337-1351, 2001.

[36] A. Bondavalli and R. Filippini, "Modeling and Analysis of a Scheduled Maintenance System: a DSPN Approach," *The Computer Journal*, 47(6), pp. 634-650, 2004.

[37] P. Lollini, *On the modelling and solution of complex systems: from two domain-specific case-studies towards the definition of a more general framework,* PhD, University of Florence, Italy, 2005.

[38] M. Kalyanakrishnan, R. K. Iyer and J. U. Patel, "Reliability of Internet Hosts: a Case Study from the End User's Perspective," *Computer Networks*, vol. 31, 47-57, 1999.

[39] D. Oppenheimer and D. A. Patterson, "Architecture and Dependability of Large-Scale Internet Services," *IEEE Internet Computing*, vol. 6, no. 5, pp. 41-49, 2002.

[40] M. Martinello, *Availability Modeling and Evaluation of Web-based Services - A pragmatic approach,* PhD, LAAS-CNRS, 2005.

[41] M. Kaâniche, K. Kanoun and M. Martinello, "A User-Perceived Availability Evaluation of a Web Based Travel Agency," *IEEE International Conference on Dependable Systems and Networks (DSN-2003),* 2003, pp. 709-718.

[42] M. Martinello, M. Kaâniche and K. Kanoun, "Web Service Availability — Impact of Error Recovery and Traffic Model," *Reliability Engineering and System Safety*, vol. 89, pp. 6-16, 2005.

[43] L. Gönczy, S. Chiaradonna, F. Di Giandomenico, A. Pataricza, A. Bondavalli and T. Bartha, "Dependability evaluation of web service-based processes," *European Performance Engineering Workshop (EPEW 2006),* 2006, pp. 166-180, (Springer).

[44] A. Bondavalli, S. Chiaradonna, F. Di Giandomenico and I. Mura, "Dependability modeling and evaluation of multiple-phased systems using DEEM," *IEEE Transactions on Reliability*, vol. 53, no. 4, pp. 509-522, 2004.

[45] S. Porcarelli, F. Di Giandomenico and A. Bondavalli, "Analyzing Quality of Service of GPRS Network Systems from a Users Perspective," in *IEEE Symposium on Computers and Communications (ISCC02 ),* Taormina, Italy, 2002.

[46] S. Porcarelli, F. Di Giandomenico, A. Bondavalli, M. Barbera and I. Mura, "Service Level Availability Estimation of GPRS," *IEEE Transactions on Mobile Computing*, vol. 2, no. 3, pp. 233-247, 2003.
[47] P. Lollini, A. Bondavalli and F. Di Giandomenico, "Evaluation of the Impact of Congestion on Service Availability in GPRS Infrastructures," in *International Service Availability Symposium (ISAS 2005),* 2005, pp. 180-195, (Springer).

**Mohamed Kaâniche** is Chargé de Recherche at LAAS-CNRS. He is a member of the research group on Dependable Computing and Fault Tolerance. From March 1997 to February 1998, he was a Visiting Research Assistant Professor at the University of Illinois at Urbana Champaign, USA. His research activities focus on the dependability and security evaluation of fault-tolerant computing systems and critical infrastructures based on analytical modeling and experimental measurements. He has (co)authored more than 80 conference and journal papers on these subjects and has been involved in several national and European projects such as CRUTIAL, HIDENETS and ReSIST. He has served on numerous program and organization committees of international conferences. He was the Program co-chair of PRDC-2004 and Program Chair of EDCC-5.

**Paolo Lollini** received the laurea degree and the PhD degree in Computer Science from the University of Firenze, Italy, in 2001 and 2005, respectively. Since June 2006, he has been a research associate at the Computer Science Dept. of the same University. He is presently participating in several national and European funded projects, including HIDENETS, CRUTIAL, SAFEDMI, and RESIST. He has authored/co-authored papers appeared in Proceedings of International Conferences, journals and books. His current research interests include the modeling and evaluation of dependability attributes, with reference to a variety of application fields, including telecommunications systems.

**Andrea Bondavalli** is a Professor of Computer Science at the University of Florence. Previously, he was a researcher at the Italian National Research Council, working at the CNUCE Institute in Pisa, where he led the Dependable Computing Group. His research is focused on the design and validation of critical systems and infrastructures. He is an author of more than 110 refereed publications in international journals and conferences and served as program chair of the most important conferences in the area, including DCC-DSN 05, IEEE SRDS 2000, EDCC-4, LADC-3. He has been PI in many projects funded by the European Community, currently IST-2004-26979 HIDENETS, and SUSTDEV-2005-2005-31413 SAFEDMI. Professor Bondavalli is a member of the IFIP W.G. 10.4 on "Dependable Computing and Fault-Tolerance.

**Karama Kanoun** is currently Directeur de Recherche at LAAS-CNRS. She was Visiting Professor at the University of Illinois, Urbana Champaign, USA, for a semester, in 1998. Her current research interests include modeling, evaluation and benchmarking of computer system dependability. She has authored or co-authored more than 120 conference and journal papers, 5 books and 10 book chapters. She has been involved in more than 30 national research contracts and European projects. She has been a consultant for several French companies, the European Space Agency, Ansaldo Transporti and for the International Union of Telecommunications. She is Chairperson of the Special Interest Group on Dependability Benchmarking (SIGDeB) of the IFIP WG 10.4 and Chairperson of the French SEE Working Group "Design and Validation for Dependability". She has co-directed the production of a book on Dependability Benchmarking, accepted for publication by IEEE Computer Society, to appear early 2008.